\documentclass[preprint,12pt,authoryear]{elsarticle}

\usepackage{amsmath,amssymb,amsthm,mathtools}
\usepackage{bm}
\usepackage{booktabs}
\usepackage{array}
\usepackage{graphicx}
\usepackage{subcaption}
\usepackage[section]{placeins}
\usepackage{enumitem}
\usepackage{microtype}
\usepackage{siunitx}
\usepackage{lineno}
\usepackage[hidelinks]{hyperref}
\usepackage[capitalize,noabbrev]{cleveref}

\graphicspath{{figures/}}
\modulolinenumbers[5]
\newtheorem{theorem}{Theorem}[section]
\newtheorem{proposition}[theorem]{Proposition}
\newtheorem{lemma}[theorem]{Lemma}
\newtheorem{corollary}[theorem]{Corollary}
\newtheorem{definition}[theorem]{Definition}
\newtheorem{remark}[theorem]{Remark}

\newcommand{\R}{\mathbb{R}}
\newcommand{\E}{\mathbb{E}}
\newcommand{\Pp}{\mathbb{P}}
\newcommand{\ind}{\mathbf{1}}
\newcommand{\dd}{\,\mathrm{d}}
\newcommand{\cL}{\mathcal{L}}

\newcommand{\pos}[1]{\left[#1\right]_{+}}
\DeclareMathOperator*{\argmin}{arg\,min}

\journal{Spatial Statistics}

\begin{document}

\begin{frontmatter}

\title{Line-constrained nearest-neighbor distance distributions and two-turn bounds for Poisson line Cox processes}

\author[aff1]{Gourab Ghatak\corref{cor1}}
\ead{gghatak@ee.iitd.ac.in}
\author[aff2]{Sanjoy Kumar Jhawar}
\ead{sanjoy-kumar.jhawar@inria.fr}
\author[aff3]{Martin Haenggi}
\ead{mhaenggi@nd.edu}

\cortext[cor1]{Corresponding author.}

\address[aff1]{Department of Electrical Engineering, Indian Institute of Technology Delhi, Hauz Khas, New Delhi 110016, India}
\address[aff2]{DYOGENE team, Inria Paris, Paris, France}
\address[aff3]{Department of Electrical Engineering, University of Notre Dame, Notre Dame, IN 46556, USA}

\begin{abstract}
Point patterns observed on streets and other linear infrastructure are often summarized using Euclidean distances even when accessibility is governed by travel along the streets. We study line-constrained nearest-neighbor and contact distributions in a Poisson line Cox process, in which an isotropic Poisson line process generates a random set of lines and independent one-dimensional Poisson processes generate points on the lines. For paths with at most one turn, we obtain a closed-form nearest-neighbor distribution under the Palm distribution of the typical Cox-process point. For the typical line intersection, we derive an exact probability-generating-functional representation for the path-length distribution that accounts for overlap between the portions of a residual line reachable from the two incident lines and uses the sine-weighted vertex-Palm angle law. We then treat paths with at most two turns without factorizing dependent branches. The exact law of a nearest-gateway restricted reachable set gives a rigorous lower bound, while the reduced-Palm Euclidean nearest-neighbor law gives a rigorous upper bound. Furthermore, we conduct Rao-Blackwellized Monte Carlo simulations to validate the bound and quantify its tightness. The characterization also yields an explicit hazard function that identifies both model intensities, leading to composite-likelihood and minimum-contrast estimation procedures. From an application perspective, we employ real data from OpenStreetMaps to calibrate the model parameters so as to map the line-length density and linear point intensity to service guarantees. The resulting summaries quantify the structural bias introduced by planar Poisson surrogates and provide controlled benchmarks for multi-turn path-lengths.
\end{abstract}


\end{frontmatter}

\section{Introduction}\label{sec:introduction}

Spatial point patterns supported on road, river, rail, fiber, and neural networks require distances and summary statistics that respect the underlying structure of the domain (e.g., the set of lines in a realization of a line process). The $K$-function and its geometrically corrected variants, path-length nearest-neighbor and empty-space functions, and inhomogeneous extensions form a substantial statistical toolkit for a fixed/observed linear set \citep{okabe2001networkK,ang2012networkK,cronie2020networksummary,baddeley2021networkreview}. A central aspect of this literature is that the choice of distance is part of the statistical model, i.e., replacing a shortest-path metric by Euclidean distance can alter both interpretation and inference.

A doubly stochastic setting arises when the domain itself is random. Random-line and random-tessellation models are useful when the objective is to characterize uncertainty across domain realizations, extrapolate beyond a single observed map, or obtain analytically tractable structural benchmarks \citep{schneider2008stochastic,chiu2013stochastic}. In a Poisson line Cox process (PLCP), an isotropic Poisson line process (PLP) generates the supporting domain and, conditionally on that domain, independent one-dimensional Poisson point processes (PPPs) generate events on the lines. Such models have been used for telecommunication access networks, vehicular systems, and transport infrastructure \citep{gloaguen2006fitting,gloaguen2010analysis,dhillon2020poisson}. They also provide a basic model against which more elaborate Cox processes on linear structures can be compared \citep{moller2024coxnetwork}.

For a planar point process, the ordinary nearest-neighbor distribution is based on Euclidean distance from a typical point. On a line process, the operational analogue is the shortest distance measured along the lines (i.e., the shortest path length). Even for a homogeneous PLCP, its unrestricted distribution is challenging to derive since a candidate route may traverse an unbounded number of line intersections. The Manhattan line Cox process (MLCP), whose line orientations are restricted to two orthogonal directions, is more tractable \citep{chetlur2020shortest}; the isotropic PLP is harder because both intersection locations and line orientations are random. Existing work has obtained mean or asymptotic shortest-path results \citep{gloaguen2010analysis,voss2010scaling}, but exact finite-distance distributions remain scarce.

This paper develops tractable first-order spatial summaries for one- and two-turn accessibility. The exact one-turn distributions define interpretable local-accessibility functionals: the nearest event that can be reached on the incident line, or after transferring once to an intersecting line. For two turns, the main difficulty is dependence: a residual line may be reached through several first-turn branches, so multiplying branchwise void probabilities is generally invalid. We avoid that factorization by analyzing one complete branch selected through the nearest line intersection and by pairing the resulting lower bound with a Euclidean upper bound. These summaries are useful for last-mile access, local dispatch, and short-range interactions, while also providing controlled benchmarks for unrestricted routing algorithms.

The contributions are as follows.
\begin{enumerate}[label=(\roman*),leftmargin=*]
    \item Under the reduced Palm distribution of a typical PLCP point, we derive a closed-form one-turn nearest-neighbor distribution, its density, and its hazard function. Then under the Palm distribution of a typical PLP intersection, we derive an exact integral representation that accounts for the union of two reachable intervals on every residual line. Furthermore, we establish computable contact-distribution bounds for the typical-intersection setting.
    \item For paths with at most two turns, we derive an exact probability-generating-functional law of the path-length for a nearest-gateway restriction. It is a rigorous lower bound on the full two-turn CDF; a one-dimensional Euclidean Palm law supplies a rigorous upper bound. Rao-Blackwellized simulation validates the sandwich.
    \item We interpret the typical-point law as a parametric nearest-neighbor summary, prove parameter identifiability, and formulate composite-likelihood and minimum-contrast estimators with an edge-aware model-checking workflow.
    \item We give an OpenStreetMap-based calibration and urban-accessibility illustration, and prove that a matched-intensity planar Poisson surrogate crosses the one-turn path-length distribution exactly once.
\end{enumerate}

The remainder of the paper is organized as follows. \Cref{sec:model} defines the PLP, the Cox process, and the turn-restricted distances. \Cref{sec:typical-point} treats the typical PLCP point, while \Cref{sec:typical-intersection} treats the typical PLP intersection. The corrected two-turn sandwich is developed in \Cref{sec:two-turn}. Statistical estimation and model checking are discussed in \Cref{sec:inference}. \Cref{sec:numerical} validates the exact laws and the two-turn bounds, and \Cref{sec:urban} develops the urban-accessibility illustration. Limitations and extensions are discussed in \Cref{sec:discussion}.

\section{Model, Palm settings, and path-length summaries}\label{sec:model}

\subsection{Poisson line Cox process}

A line is represented by a direction angle $\alpha\in[0,\pi)$ and a signed perpendicular displacement $p\in\R$:
\begin{align}
L(p,\alpha)
&=\left\{\bm{x}\in\R^2:\bm{n}_{\alpha}^{\mathsf T}\bm{x}=p\right\}, \;\; \bm{n}_{\alpha}=(-\sin\alpha,\cos\alpha)^T
\label{eq:line-parameterization}
\end{align}
The vector $\bm{u}_{\alpha}=(\cos\alpha,\sin\alpha)^T,$ gives the line direction and $\bm{n}_{\alpha}$ its unit normal.

\begin{definition}[Poisson line process]
Let $\Xi$ be a PPP on $\R\times[0,\pi)$ with intensity measure
\begin{equation}
\Lambda_{\Xi}(\dd p,\dd\alpha)=\lambda\,\dd p\,\dd\alpha,
\qquad \lambda>0.
\label{eq:line-intensity}
\end{equation}
The random collection $\cL=\{L(p,\alpha):(p,\alpha)\in\Xi\}$ is an isotropic stationary Poisson line process.
\end{definition}

Under this convention, $\lambda$ has dimension inverse length and
\begin{equation}
\E\!\left[\operatorname{length}(\cL\cap W)\right]
=\pi\lambda |W|
\label{eq:line-length-density}
\end{equation}
for every bounded Borel window $W\subset\R^2$. Thus, the mean total line length per unit area is $\pi\lambda$. Conditionally on $\cL$, let us place an independent stationary one-dimensional PPP of intensity $\mu>0$ on every line. Their union is the PLCP $\Phi$. Using \eqref{eq:line-length-density}, its planar intensity is
\begin{equation}
\rho=\pi\lambda\mu.
\label{eq:planar-intensity}
\end{equation}
The two parameters have distinct interpretations: $\lambda$ controls the random line density, whereas $\mu$ controls the event density along the lines.

\subsection{Path-Lengths and the two Palm origins}

For points $\bm{x},\bm{y} \in \Phi$, let $d_{\cL}(\bm{x},\bm{y})$ denote the geodesic distance along the union of lines. We use $d_{\cL}^{(k)}$ for the corresponding distance when paths are restricted to at most $k$ changes of supporting line. More precisely, the superscript~$k$ specifies an upper bound on the number of changes of supporting line, rather than an exact number of changes. For example, a path that remains on an initial line is admissible for every $k\geq 0$, while $d_{\mathcal L}^{(1)}$ permits either no change of line or one transfer to a
second line. Note that unlike a Manhattan line process, the line orientations are not restricted to the coordinate axes\footnote{Hence calling the geodesics ``$\ell_1$-distances'' maybe inaccurate.}.

We consider two origin-sampling schemes.
\begin{enumerate}[label=(\alph*),leftmargin=*]
    \item \emph{Typical PLCP point.} Under the reduced Palm distribution of $\Phi$, place the typical point at the origin and remove it from the process. An independent Palm line passes through the origin; isotropy allows it to be taken as the $x$-axis, denoted $L_0$.
    \item \emph{Typical PLP intersection.} Under the Palm distribution of the intersection process of $\cL$, place the typical vertex at the origin. Two Palm lines $L_0$ and $L_{\Theta}$ pass through the origin, and the residual PLP is independent of them.
\end{enumerate}

The principal distance variables are
\begin{align}
D_{\mathrm P}
&=\inf_{\bm{x}\in\Phi\setminus\{\bm{0}\}}
 d_{\cL}^{(1)}(\bm{0},\bm{x}),
\label{eq:Dpoint}\\
D_2
&=\inf_{\bm{x}\in\Phi\setminus\{\bm{0}\}}
 d_{\cL}^{(2)}(\bm{0},\bm{x}),
\label{eq:Dtwo}\\
D_{\mathrm I}
&=\inf_{\bm{x}\in\Phi}
 d_{\cL}^{(1)}(\bm{0},\bm{x}).
\label{eq:Dintersection}
\end{align}
The first two are path-length nearest-neighbor distances under a reduced Palm distribution, with at most one and two turns, respectively. The third is a contact distance from the typical intersection of the line process; the origin need not be a point of $\Phi$.

\begin{figure}[htbp]
    \centering
    \includegraphics[width=0.72\textwidth]{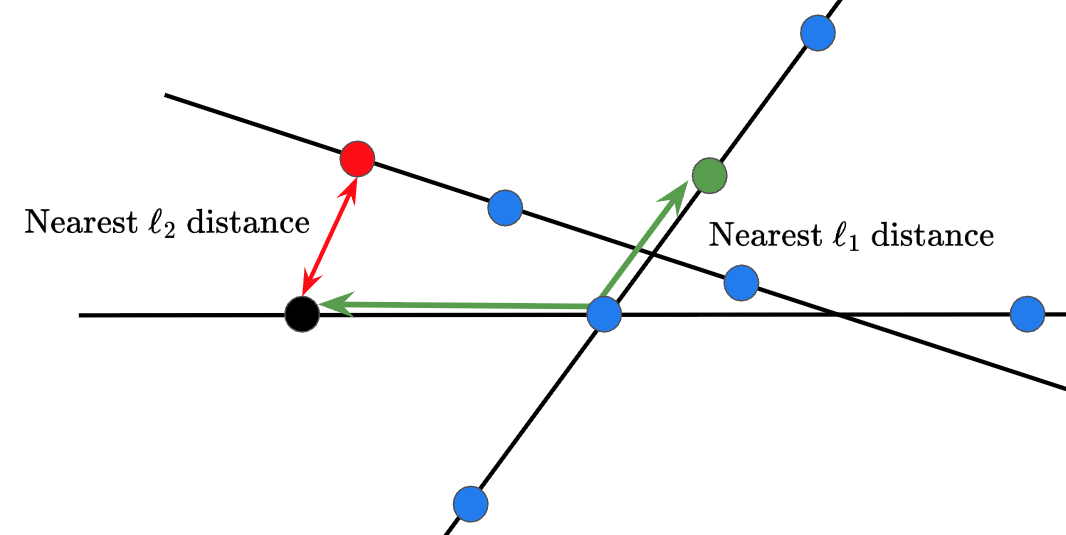}
    \caption{Euclidean and line-constrained proximity need not select the same point. From the black origin, the red point is nearest in Euclidean distance, whereas the green point is nearest under the street-constrained path metric.}
    \label{fig:distance-illustration}
\end{figure}

\begin{figure}[htbp]
    \centering
    \begin{subfigure}[t]{0.46\textwidth}
        \centering
        \includegraphics[width=\linewidth]{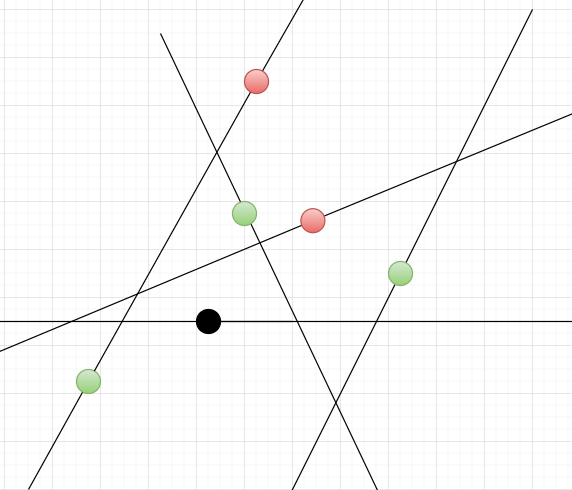}
        \caption{Typical PLCP point.}
        \label{fig:typical-point-geometry}
    \end{subfigure}\hfill
    \begin{subfigure}[t]{0.46\textwidth}
        \centering
        \includegraphics[width=\linewidth]{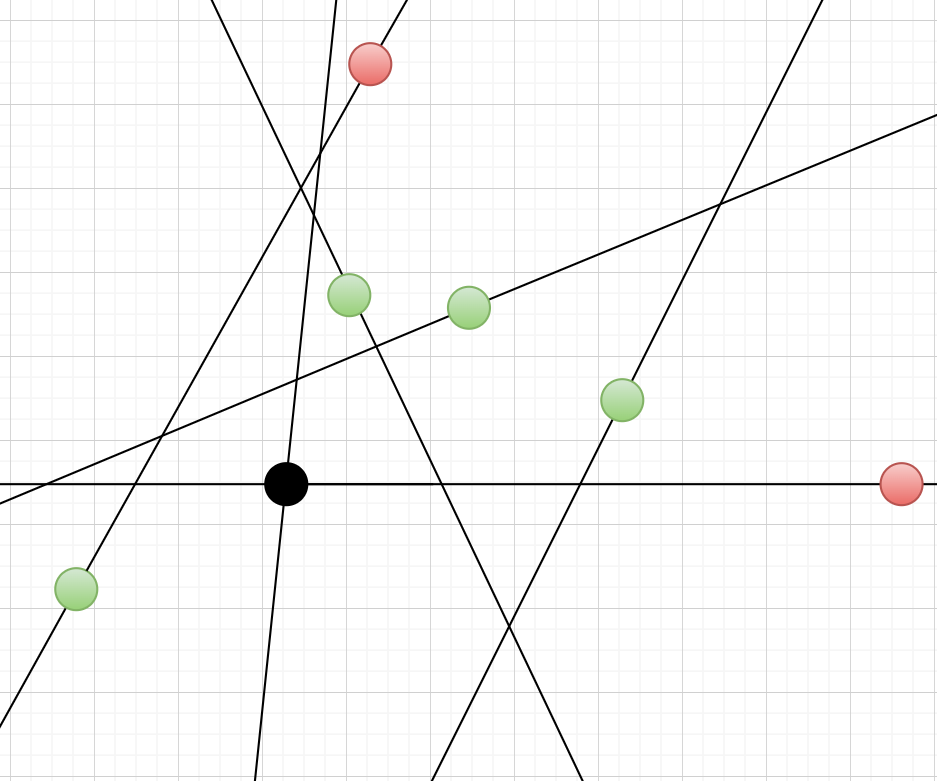}
        \caption{Typical PLP intersection.}
        \label{fig:typical-intersection-geometry}
    \end{subfigure}
    \caption{One-turn reachable sets within path-length budget $t$. Green points are reachable from the black origin with at most one transfer.}
    \label{fig:one-turn-geometries}
\end{figure}

\begin{table}[htbp]
\centering
\caption{Field-specific terms and principal notation.}
\label{tab:notation}
\begin{tabular}{>{\raggedright\arraybackslash}p{0.22\textwidth}p{0.68\textwidth}}
\toprule
Term or symbol & Definition \\
\midrule
PLP, $\cL$ & Poisson process of infinite lines generated in Hesse space by \eqref{eq:line-intensity}. \\
PLCP, $\Phi$ & Cox point process obtained by placing independent one-dimensional Poisson processes on the PLP lines. \\
$\lambda$ & Hesse-space line intensity; the mean line-length density is $\pi\lambda$. \\
$\mu$ & Conditional point intensity per unit line length. \\
$d_{\cL}^{(k)}$ & Shortest path-length among paths making at most $k$ changes of supporting line. \\
$D_2$ & At-most-two-turn nearest-neighbor distance from a typical PLCP point. \\
$D_{\mathrm P}$ & One-turn nearest-neighbor distance from a typical PLCP point. \\
$D_{\mathrm I}$ & One-turn contact distance from a typical PLP intersection. \\
$D_{\mathrm G}$ & Nearest-gateway restricted distance used to lower-bound $D_2$. \\
$F_{\mathrm P},F_{\mathrm I}$ & Distribution functions of $D_{\mathrm P}$ and $D_{\mathrm I}$. \\
\bottomrule
\end{tabular}
\end{table}

\section{Typical-point nearest-neighbor distribution}\label{sec:typical-point}

Under the reduced Palm distribution of $\Phi$, let $L_0$ be the line through the origin. The intersections of the residual PLP with $L_0$ form a stationary Poisson process on $L_0$ with intensity $2\lambda$. Indeed, for a residual line $L(p,\alpha)$, its signed intersection coordinate on $L_0$ is $s=-p/\sin\alpha$, and the mapping theorem gives
\begin{equation}
\lambda\int_0^{\pi}|\sin\alpha|\dd\alpha=2\lambda.
\label{eq:intersection-rate}
\end{equation}

\begin{theorem}[Typical-point one-turn law]\label{thm:typical-point-law}
For $t\geq0$, the survival function and distribution function of $D_{\mathrm P}$ are
\begin{align}
S_{\mathrm P}(t)
&=\Pp(D_{\mathrm P}>t)\nonumber\\
&=\exp\!\left\{-2\mu t-4\lambda t
+\frac{2\lambda}{\mu}\bigl(1-\exp(-2\mu t)\bigr)\right\},
\label{eq:SP}\\
F_{\mathrm P}(t)&=1-S_{\mathrm P}(t).
\label{eq:FP}
\end{align}
\end{theorem}

\begin{proof}
The direct portion of $L_0$ reachable within distance $t$ has length $2t$, and its conditional void probability is $\exp(-2\mu t)$. A residual line intersecting $L_0$ at signed coordinate $s\in[-t,t]$ contributes a reachable segment of length $2(t-|s|)$, whose conditional void probability is $\exp\{-2\mu(t-|s|)\}$. Applying the probability generating functional of the intersection process in \eqref{eq:intersection-rate},
\begin{align}
S_{\mathrm P}(t)
&=\exp(-2\mu t)
\exp\!\left[-2\lambda\int_{-t}^{t}
\left\{1-\exp\bigl(-2\mu(t-|s|)\bigr)\right\}\dd s\right]\\
&=\exp(-2\mu t)
\exp\!\left[-4\lambda\int_0^t
\left\{1-\exp\bigl(-2\mu(t-s)\bigr)\right\}\dd s\right],
\end{align}
which evaluates to \eqref{eq:SP}.
\end{proof}

\begin{corollary}[Density and hazard]\label{cor:hazard}
The density of $D_{\mathrm P}$ is $f_{\mathrm P}(t)=h_{\mathrm P}(t)S_{\mathrm P}(t)$, where
\begin{equation}
h_{\mathrm P}(t)
=2\mu+4\lambda\bigl(1-\exp(-2\mu t)\bigr).
\label{eq:hazard}
\end{equation}
Consequently,
\begin{equation}
h_{\mathrm P}(0)=2\mu,
\qquad
\lim_{t\to\infty}h_{\mathrm P}(t)=2\mu+4\lambda,
\label{eq:hazard-limits}
\end{equation}
so the immediate hazard is determined by points on the Palm line. At large distances, the expression reflects the additional opportunities created by intersecting lines.
\end{corollary}


\section{Contact distribution from the typical line intersection}\label{sec:typical-intersection}

The typical-intersection setting differs from the typical-point setting in two ways. First, two initial lines are available. Second, a residual line can be reached from either initial line, so the two reachable intervals on that residual line may overlap. Treating the two intervals as independent would double-count their overlap.

\begin{figure}
    \centering
    \includegraphics[width=0.8\linewidth]{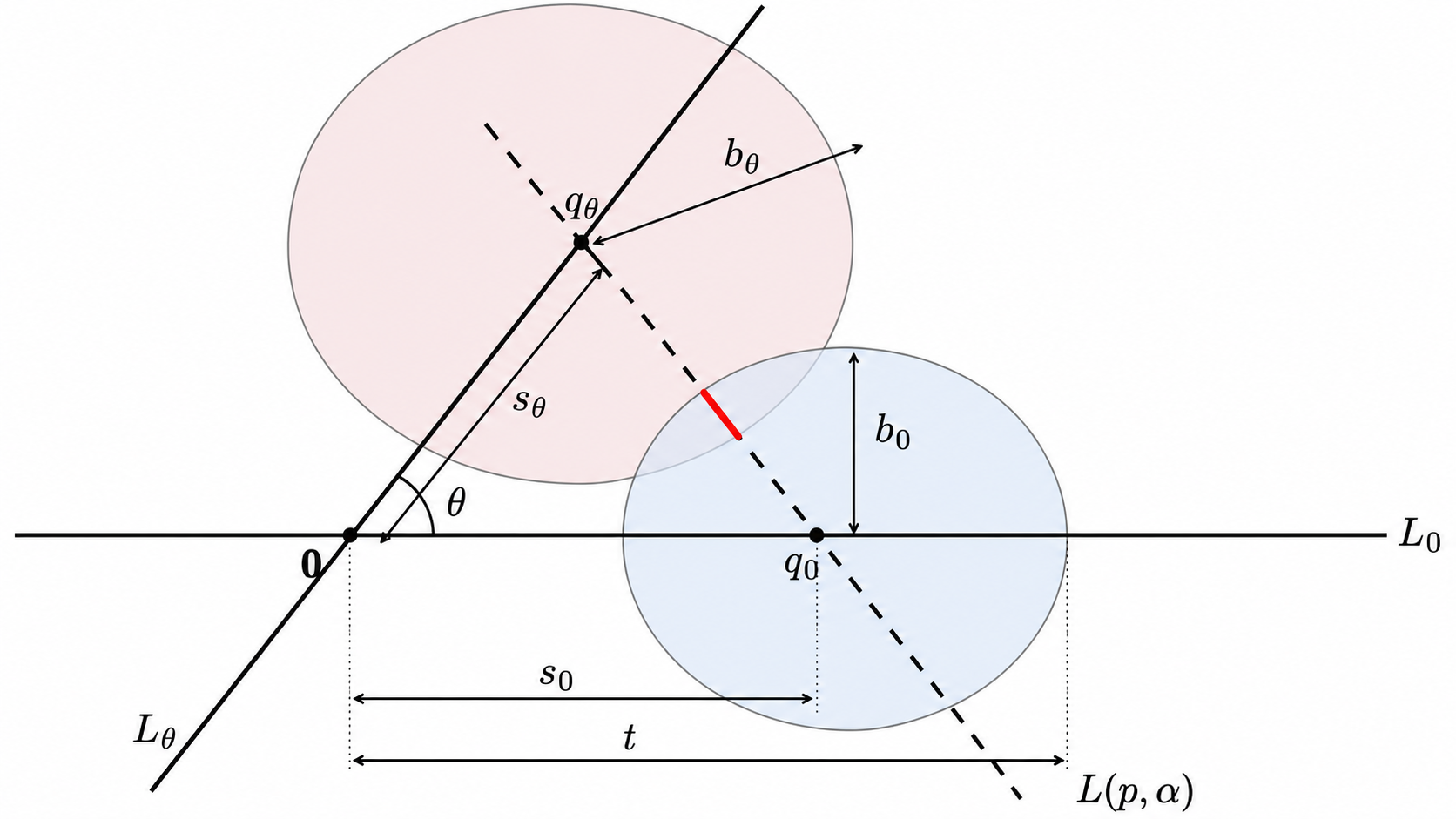}
    \caption{The single turn case starting from the typical intersection.}
    \label{fig:Intersection}
\end{figure}

\subsection{Palm distribution of the intersection angle}

\begin{lemma}[Typical-intersection angle]\label{lem:angle-law}
Let $\Theta\in(0,\pi)$ be the angle between the two lines forming the typical PLP intersection. Then
\begin{equation}
f_{\Theta}(\theta)=\frac{1}{2}\sin\theta,
\qquad 0<\theta<\pi.
\label{eq:angle-density}
\end{equation}
Equivalently, the acute angle $A=\min(\Theta,\pi-\Theta)$ has density $f_A(a)=\sin a$ on $(0,\pi/2)$.
\end{lemma}

\begin{proof}
For two lines with direction angles $\alpha_1$ and $\alpha_2$, the change of variables from their signed offsets to their intersection location has Jacobian $|\sin(\alpha_2-\alpha_1)|$. Hence, under vertex Palm sampling, pairs with angle difference $\theta$ receive weight proportional to $\sin\theta$. Normalization by $\int_0^{\pi}\sin u\dd u=2$ gives \eqref{eq:angle-density}; see also standard intersection-process constructions in stochastic and integral geometry \citep{schneider2008stochastic,chiu2013stochastic}.
\end{proof}

\subsection{Reachable intervals on a residual line}

By isotropy, condition on $L_0$ having direction $0$ and $L_{\theta}$ having direction $\theta$. Consider a residual line $L(p,\alpha)$. Its signed coordinates along the two initial lines are
\begin{equation}
s_0=-\frac{p}{\sin\alpha},
\qquad
s_{\theta}=\frac{p}{\sin(\theta-\alpha)},
\label{eq:initial-line-coordinates}
\end{equation}
and the corresponding coordinates along the residual line are
\begin{equation}
q_0=-p\cot\alpha,
\qquad
q_{\theta}=p\cot(\theta-\alpha).
\label{eq:residual-line-coordinates}
\end{equation}
The remaining path-length budgets after reaching the residual line are
\begin{equation}
b_0=(t-|s_0|)_+,
\qquad
b_{\theta}=(t-|s_{\theta}|)_+,
\label{eq:remaining-budgets}
\end{equation}
where $(x)_+=\max(x,0)$. The reachable intervals, expressed in the coordinate of $L(p,\alpha)$, are
\begin{equation}
I_0=[q_0-b_0,q_0+b_0],
\qquad
I_{\theta}=[q_{\theta}-b_{\theta},q_{\theta}+b_{\theta}],
\end{equation}
with an interval regarded as empty when the corresponding budget is zero. Their union length is
\begin{align}
\ell_t(p,\alpha;\theta)
={}&2b_0+2b_{\theta}\nonumber\\
&-\pos{\min(q_0+b_0,q_{\theta}+b_{\theta})
-\max(q_0-b_0,q_{\theta}-b_{\theta})}.
\label{eq:union-length}
\end{align}
The formula is interpreted by continuity at the measure-zero orientations for which a denominator in \eqref{eq:initial-line-coordinates} vanishes. Moreover, $\ell_t=0$ unless
\begin{equation}
|p|\leq a_t(\alpha,\theta)
\coloneq t\max\{|\sin\alpha|,|\sin(\theta-\alpha)|\}.
\label{eq:p-support}
\end{equation}

\begin{theorem}[Typical-intersection one-turn law]\label{thm:intersection-law}
Define
\begin{equation}
H(t,\theta)
=\int_0^{\pi}\int_{-a_t(\alpha,\theta)}^{a_t(\alpha,\theta)}
\left[1-\exp\{-\mu\ell_t(p,\alpha;\theta)\}\right]
\dd p\dd\alpha.
\label{eq:H-function}
\end{equation}
Then the survival and distribution functions of $D_{\mathrm I}$ are
\begin{align}
S_{\mathrm I}(t)
&=\frac{1}{2}\int_0^{\pi}
\exp\{-4\mu t-\lambda H(t,\theta)\}
\sin\theta\dd\theta,
\label{eq:SI}\\
F_{\mathrm I}(t)&=1-S_{\mathrm I}(t).
\label{eq:FI}
\end{align}
\end{theorem}

\begin{proof}
Condition on $\Theta=\theta$ and on the residual PLP. The total length directly reachable on $L_0\cup L_{\theta}$ is $4t$, giving the factor $\exp(-4\mu t)$. On a residual line $L(p,\alpha)$, the set of points reachable with one turn is exactly $I_0\cup I_{\theta}$, whose conditional void probability is $\exp\{-\mu\ell_t(p,\alpha;\theta)\}$. The probability generating functional of the residual line process therefore yields
\begin{equation}
\Pp(D_{\mathrm I}>t\mid\Theta=\theta)
=\exp\{-4\mu t-\lambda H(t,\theta)\}.
\end{equation}
Averaging this conditional probability with respect to \eqref{eq:angle-density} gives \eqref{eq:SI}.
\end{proof}

The expression in \eqref{eq:SI} is a three-dimensional bounded integral. It avoids a case-by-case partition of residual-line orientations. For numerical evaluation, we use nested Gauss--Legendre quadrature on fixed
compact intervals. In particular, for fixed $(t,\theta,\alpha)$, the change
of variable
\[
    p=a_t(\alpha,\theta)z, \qquad -1\leq z\leq 1,
\]
rewrites~\eqref{eq:H-function} as
\[
H(t,\theta)
 = \int_0^\pi a_t(\alpha,\theta)
   \int_{-1}^{1}
   \left[
   1-\exp\!\left\{
      -\mu\ell_t\bigl(a_t(\alpha,\theta)z,\alpha;\theta\bigr)
   \right\}
   \right]
   \,\mathrm dz\,\mathrm d\alpha .
\]
The $\alpha$-integral and the outer $\theta$-average
in~\eqref{eq:SI} are then mapped affinely from $[-1,1]$ and evaluated
with tensor-product Gauss--Legendre rules. This transformation keeps every
numerical integration bounded and avoids truncation of an infinite domain. Importantly, the average over the Palm angle occurs \emph{outside} the exponential because the angle is common to the full residual-line configuration.

\subsection{Bounds}

\begin{proposition}[Contact-distribution bounds]\label{prop:intersection-bounds}
For $t\geq0$,
\begin{align}
1-\exp(-4\mu t)
\leq F_{\mathrm I}(t)
\leq{}&1-\exp(-4\mu t)\frac{1}{2}
\int_0^{\pi}\exp\{-\lambda M(t,\theta)\}\sin\theta\dd\theta,
\label{eq:intersection-bounds}
\end{align}
where
\begin{equation}
M(t,\theta)
=4t\left\{\sin(\theta/2)+\cos(\theta/2)\right\}.
\label{eq:Mtheta}
\end{equation}
Consequently,
\begin{equation}
F_{\mathrm I}(t)
\leq 1-\exp\{-(4\mu+4\sqrt{2}\lambda)t\}.
\label{eq:simple-upper-bound}
\end{equation}
\end{proposition}

\begin{proof}
Using only the two initial line segments gives a subset of the reachable set, so their void probability $\exp(-4\mu t)$ upper-bounds the full void probability, yielding the lower bound on $F_{\mathrm I}$.

For the upper bound, consider the sufficient event that (i) the two initial segments contain no PLCP points and (ii) no residual PLP line intersects either segment. For direction $\alpha$, the set of signed offsets of lines intersecting the union of the two segments is an interval of length
\begin{equation}
2t\max\{|\sin\alpha|,|\sin(\theta-\alpha)|\}.
\end{equation}
Integration over $\alpha$ gives
\begin{align}
M(t,\theta)
&=2t\int_0^{\pi}\max\{|\sin\alpha|,|\sin(\theta-\alpha)|\}\dd\alpha\\
&=4t\{\sin(\theta/2)+\cos(\theta/2)\}.
\end{align}
The probability of condition (ii), given $\Theta=\theta$, is $\exp\{-\lambda M(t,\theta)\}$. Averaging the resulting lower bound on the survival probability gives \eqref{eq:intersection-bounds}. Finally, $M(t,\theta)\leq4\sqrt{2}t$ gives \eqref{eq:simple-upper-bound}.
\end{proof}

The lower bound retains only the two initial Palm-line
segments and is therefore most informative when direct access dominates, for
example when $\lambda t$ is small. It becomes exact in the degenerate case in
which no residual line contributes to the reachable set. The upper bound
additionally requires that no residual line intersect either initial segment.
It is consequently conservative because a residual line may intersect the
segments and nevertheless contain no Cox-process point in its reachable
portion; this loss is less pronounced when $\mu t$ is large. The exponential
bound is simpler still, at the cost of replacing
the angle-dependent exclusion measure by its worst-case value.

These inequalities provide a low-cost screening rule for service design. For
a target distance $d_0$ and reliability level $\eta$, the condition
\[
    1-\exp(-4\mu d_0)\geq \eta
\]
already certifies
\[
    \mathbb P(D_{\mathrm I}\leq d_0)\geq\eta.
\]
Conversely, if the upper bound is smaller than $\eta$, then the target cannot be met for the specified $(\lambda,\mu)$. Only the unresolved intermediate
regime requires evaluation of the exact integral in~\eqref{eq:SI}. This screening interpretation is used again in the service-guarantee discussion of
Section~8.2.
\section{Bounds for paths with at most two turns}\label{sec:two-turn}

The one-turn law in \Cref{thm:typical-point-law} is exact, but allowing a second transfer creates dependencies because the same residual line may be reached through several first-turn lines. Rather than factorizing these dependent branches, we construct an exactly analysable subset of the two-turn reachable set and use it to obtain a rigorous lower bound. A complementary upper bound follows from Euclidean proximity. Together they bracket the full two-turn distribution without an independence approximation.

Recall $D_2$ from \eqref{eq:Dtwo}, and write $F_2(t)=\Pp(D_2\leq t)$. The exact distribution of $D_2$ is not required for the bounds below.

\subsection{Nearest-gateway restriction}

Write $(X_i,A_i)$ for the signed intersection coordinate and direction mark of residual line $i$ on the Palm line $L_0$. Recall that by the mapping used in \eqref{eq:intersection-rate}, these pairs form a Poisson process on $\R\times(0,\pi)$ with intensity
\begin{equation}
\lambda\sin a\,\dd x\dd a.
\label{eq:marked-intersection-intensity}
\end{equation}
Integrating the joint intensity in~\eqref{eq:marked-intersection-intensity} over the
angle mark gives
\(
    \int_0^\pi \lambda\sin a\,\mathrm da = 2\lambda,
\)
which recovers the signed-intersection intensity
in~\eqref{eq:intersection-rate}. Equivalently,
\(    \lambda\sin a\,\mathrm dx\,\mathrm da
    =
    \bigl(2\lambda\,\mathrm dx\bigr)
    \left(\frac{1}{2}\sin a\,\mathrm da\right).
\)
Hence the signed coordinates form a stationary Poisson process of rate
$2\lambda$, and, conditionally on an intersection coordinate, the independent
direction mark has density
\[
    f_A(a)=\frac{1}{2}\sin a,
    \qquad 0<a<\pi.
\]

Let
\begin{equation}
S=\min_i |X_i|
\label{eq:nearest-gateway-distance}
\end{equation}
be the distance to the nearest line intersection on either side of the origin, and call the corresponding line the \emph{gateway line}. The nearest-gateway reachable set contains: (i) every point reachable with at most one turn; and (ii) every point reachable with two turns when the first transfer is made to the gateway line. Let $D_{\mathrm G}$ and $F_{\mathrm G}$ denote the associated nearest-point distance and distribution function. This construction preserves all one-turn opportunities while retaining one complete two-turn branch.

Condition on $S=s\leq t$ and, by reflection symmetry, place the gateway intersection at $(s,0)$. Let its direction be $\gamma$. For another residual line $L(p,\alpha)$, use the coordinate representation
\begin{equation}
L(p,\alpha)=\{p\bm{n}_{\alpha}+q\bm{u}_{\alpha}:q\in\R\}.
\end{equation}
The intersections with $L_0$ and with the gateway line have line coordinates and remaining path-length budgets
\begin{subequations}\label{eq:gateway-coordinates}
\begin{align}
q_0(p,\alpha)
&=-p\cot\alpha,\\
b_0(p,\alpha)
&=\pos{t-\frac{|p|}{\sin\alpha}},\\
v_{\mathrm G}(p,\alpha;s,\gamma)
&=\frac{p+s\sin\alpha}{\sin(\gamma-\alpha)},\\
b_{\mathrm G}(p,\alpha;s,\gamma)
&=\pos{t-s-|v_{\mathrm G}(p,\alpha;s,\gamma)|},\\
q_{\mathrm G}(p,\alpha;s,\gamma)
&=s\cos\alpha+v_{\mathrm G}(p,\alpha;s,\gamma)
 \cos(\gamma-\alpha).
\end{align}
\end{subequations}
The measure-zero parallel cases are interpreted by continuity, with the corresponding interval empty when no intersection is reachable. Furthermore, define
\begin{align}
I_0&=[q_0-b_0,q_0+b_0],\nonumber\\
I_{\mathrm G}&=[q_{\mathrm G}-b_{\mathrm G},q_{\mathrm G}+b_{\mathrm G}],\nonumber\\
\ell_{\mathrm G}(t;p,\alpha,s,\gamma)
&=\operatorname{length}(I_0\cup I_{\mathrm G}).
\label{eq:gateway-union-length}
\end{align}
Equivalently,
\begin{align}
\ell_{\mathrm G}
={}&2b_0+2b_{\mathrm G}\nonumber\\
&-\pos{\min(q_0+b_0,q_{\mathrm G}+b_{\mathrm G})
-\max(q_0-b_0,q_{\mathrm G}-b_{\mathrm G})}.
\label{eq:gateway-union-explicit}
\end{align}
The conditioning $S=s$ excludes every other line whose intersection with $L_0$ lies in $(-s,s)$. In Hesse coordinates this is the region $|p|<s\sin\alpha$. Hence define the finite integral
\begin{align}
K_{\mathrm G}(t;s,\gamma)
={}&\int_0^{\pi}\int_{|p|\geq s\sin\alpha}
\left[1-\exp\{-\mu\ell_{\mathrm G}(t;p,\alpha,s,\gamma)\}\right]
\dd p\dd\alpha.
\label{eq:KG}
\end{align}
The integrand vanishes outside the union of
\begin{align}
|p|&\leq t\sin\alpha,
\label{eq:gateway-support-1}\\
|p+s\sin\alpha|
&\leq (t-s)|\sin(\gamma-\alpha)|,
\label{eq:gateway-support-2}
\end{align}
so \eqref{eq:KG} is a bounded numerical integral.

\begin{theorem}\label{thm:two-turn-sandwich}
For $t\geq0$, the survival function of the nearest-gateway restricted distance is
\begin{align}
S_{\mathrm G}(t)
={}&\exp\{-4\lambda t-2\mu t\}\nonumber\\
&+\int_0^t 4\lambda e^{-4\lambda s}
\frac{1}{2}\int_0^{\pi}
\exp\{-2\mu t-2\mu(t-s)\}\nonumber\\
&\hspace{5em}\times
\exp\{-\lambda K_{\mathrm G}(t;s,\gamma)\}
\sin\gamma\dd\gamma\dd s,
\label{eq:SG}
\end{align}
with $F_{\mathrm G}(t)=1-S_{\mathrm G}(t)$. Moreover, if $R$ is the Euclidean nearest-neighbour distance from the typical PLCP point and
\begin{align}
F_{\mathrm E}(t)
=1-\exp\left\{
-2\mu t
-2\pi\lambda\int_0^t
\left[1-\exp\{-2\mu\sqrt{t^2-p^2}\}\right]\dd p
\right\},
\label{eq:FE}
\end{align}
then
\begin{equation}
F_{\mathrm P}(t)
\leq F_{\mathrm G}(t)
\leq F_2(t)
\leq F_{\mathrm E}(t).
\label{eq:two-turn-sandwich}
\end{equation}
Thus, $F_{\mathrm G}$ is a rigorous lower bound on the full two-turn distribution and $F_{\mathrm E}$ is a rigorous upper bound.
\end{theorem}

\begin{proof}
Integrating \eqref{eq:marked-intersection-intensity} over the direction mark shows that the signed intersections form a stationary Poisson process of rate $2\lambda$. The absolute intersection distances therefore form a Poisson process on $\R_+$ of rate $4\lambda$, so
\begin{equation}
S\sim\operatorname{Exp}(4\lambda).
\label{eq:S-law}
\end{equation}
The gateway direction is independent of $S$ and has density
\begin{equation}
f_{A}(\gamma)=\frac{1}{2}\sin\gamma,
\qquad 0<\gamma<\pi.
\label{eq:gateway-angle-density}
\end{equation}
Conditioned on $S=s$ and the gateway line, Slivnyak's theorem and independent increments leave a Poisson line process on the complementary Hesse-space region $|p|\geq s\sin\alpha$.

Suppose first that $s\leq t$. The directly reachable segment of $L_0$ has length $2t$. After traveling distance $s$ to the gateway intersection, a segment of length $2(t-s)$ is reachable on the gateway line. On every remaining line $L(p,\alpha)$, the one-turn and gateway-mediated intervals are exactly $I_0$ and $I_{\mathrm G}$. Conditional independence of the one-dimensional Poisson processes on distinct lines and the probability generating functional of the residual PLP give
\begin{align}
\Pp(D_{\mathrm G}>t\mid S=s,A=\gamma)
={}&\exp\{-2\mu t-2\mu(t-s)\}
\exp\{-\lambda K_{\mathrm G}(t;s,\gamma)\}.
\label{eq:conditional-SG}
\end{align}
If $s>t$, no residual line can be reached from $L_0$ within the budget, so the restricted reachable set consists only of the length-$2t$ segment of $L_0$. Its void probability is $e^{-2\mu t}$, and the probability of $S>t$ is $e^{-4\lambda t}$. Averaging \eqref{eq:conditional-SG} using \eqref{eq:S-law}--\eqref{eq:gateway-angle-density} yields \eqref{eq:SG}.

The nearest-gateway reachable set contains the complete one-turn reachable set and is contained in the complete at-most-two-turn reachable set. Consequently,
\begin{equation}
D_2\leq D_{\mathrm G}\leq D_{\mathrm P},
\end{equation}
which proves the first two inequalities in \eqref{eq:two-turn-sandwich}.

For the final inequality, every path of length at most $t$ ends inside the Euclidean disk $B(\bm{0},t)$. Hence $R\leq D_2$ and $F_2(t)\leq\Pp(R\leq t)$. Under the reduced Palm distribution, the Palm line contributes a disk chord of length $2t$. A residual line with Hesse offset $p$ contributes a chord of length $2\sqrt{t^2-p^2}$ when $|p|\leq t$. Applying the line-process probability generating functional gives
\begin{align}
\Pp(R>t)
={}&e^{-2\mu t}
\exp\left[-\lambda\int_0^{\pi}\int_{-t}^{t}
\left\{1-e^{-2\mu\sqrt{t^2-p^2}}\right\}\dd p\dd\alpha\right],
\end{align}
which simplifies to $1-F_{\mathrm E}(t)$ in \eqref{eq:FE}.
\end{proof}

\begin{remark}\label{rem:bound-direction}
Restricting the family of admissible two-turn paths can only increase the nearest-point distance. Therefore it produces a \emph{lower} bound on the CDF, not an upper bound. The upper bound in \eqref{eq:two-turn-sandwich} comes from enlarging the candidate set from reachable points to all PLCP points in the Euclidean disk.
\end{remark}

\begin{remark}[Scale invariance]\label{rem:gateway-scaling}
Set $s=tu$ and $p=tz$, where $u\in[0,1]$. All path-length budgets and all
reachable interval lengths then scale linearly with $t$. More precisely, for
an appropriate dimensionless union length
$\bar\ell_{\mathrm G}$,
\[
    \ell_{\mathrm G}(t;tz,\alpha,tu,\gamma)
    =
    t\,\bar\ell_{\mathrm G}(z,\alpha,u,\gamma),
\]
and hence
\[
    K_{\mathrm G}(t;tu,\gamma)
    =
    t\,\bar K_{\mathrm G}(u,\gamma;\mu t).
\]
It follows from~(42) that $S_{\mathrm G}(t)$ depends on
$(\lambda,\mu,t)$ only through the two dimensionless products
$\lambda t$ and $\mu t$. Thus, a single table of dimensionless curves can be
rescaled to any physical distance unit, and parameter studies need only
explore a two-dimensional domain.
\end{remark}

\begin{remark}[Numerical evaluation]\label{rem:gateway-computation}
After the substitutions $s=tu$ and $p=tz$, the offset exclusion and support
conditions in~(39)--(41) become
\[
    |z|\geq u\sin\alpha,
\]
with a nonzero integrand only if at least one of
\[
    |z|\leq \sin\alpha,
    \qquad
    |z+u\sin\alpha|
    \leq (1-u)|\sin(\gamma-\alpha)|
\]
holds. For fixed $(u,\gamma,\alpha)$, we partition the $z$-axis at the finite
set of support and conditioning breakpoints
\[
    \pm u\sin\alpha,
    \qquad
    \pm\sin\alpha,
    \qquad
    -u\sin\alpha
    \pm(1-u)|\sin(\gamma-\alpha)|.
\]
Gauss--Legendre quadrature is then applied separately on each active
subinterval. This avoids evaluating across the kinks introduced by the
positive-part operations and by the union length
$\operatorname{length}(I_0\cup I_{\mathrm G})$. The outer variables
$u$, $\gamma$, and $\alpha$ are also integrated by Gauss--Legendre rules on
$[0,1]$ or $[0,\pi]$. For reproducibility, Section~7 reports the quadrature
orders used and the maximum change observed when all orders are doubled.
\end{remark}
\section{Statistical use: estimation and model checking}\label{sec:inference}

The functions $F_{\mathrm P}$ and $F_{\mathrm I}$ can be viewed as parametric first-order summaries of a random linear-network point process. They complement the network $K$-, pair-correlation, empty-space, nearest-neighbor, and $J$-function methodology developed for fixed networks \citep{okabe2001networkK,ang2012networkK,cronie2020networksummary,baddeley2021networkreview}. The distinction is that the expectation here integrates over both the random point pattern and the random supporting line set.

\subsection{Identifiability from the typical-point hazard}

\begin{remark}[Parameter identifiability]\label{prop:identifiability}
The mapping $(\lambda,\mu)\mapsto F_{\mathrm P}$ is identifiable. In particular,
\begin{equation}
\mu=\frac{h_{\mathrm P}(0)}{2},
\qquad
\lambda=\frac{1}{4}\left\{\lim_{t\to\infty}h_{\mathrm P}(t)-h_{\mathrm P}(0)\right\}.
\label{eq:identification}
\end{equation}
\end{remark}


This result is useful statistically because $\mu$ controls the near-origin behavior, while $\lambda$ controls the additional hazard accumulated through line intersections. The two effects are therefore not confounded in the full distance distribution.

\subsection{Composite likelihood and minimum contrast}
The identifiability result above shows that the full typical-point distance
law contains separate information about the line intensity $\lambda$ and the
along-line point intensity $\mu$. The purpose of this section is to turn that
observation into estimators based on observed turn-restricted
nearest-neighbour distances. Such distances may come from independent
replicated windows or from many interior origins in one large realization.
In the latter case they are generally spatially dependent, so the product of
their marginal densities is interpreted as a composite likelihood rather
than as an independent-sample likelihood. Minimum contrast provides a
complementary likelihood-free procedure based on matching the empirical and
model distribution functions. Both approaches target the marginal summary
$F_{\mathrm P}$, or $F_{\mathrm I}$ for vertex-origin data, while dependence
is handled through block resampling, increasing-domain arguments, or
parametric simulation.

Suppose $D_1,\ldots,D_n$ are one-turn nearest-neighbor distances extracted from replicated windows, or from sufficiently separated interior origins in a large observation window. Treating their marginal laws as the primary modeling target gives the composite log-likelihood
\begin{equation}
\ell_{\mathrm c}(\lambda,\mu)
=\sum_{i=1}^{n}w_i
\left[\log h_{\mathrm P}(D_i)+\log S_{\mathrm P}(D_i)\right],
\label{eq:composite-likelihood}
\end{equation}
where $w_i$ may downweight origins with incomplete neighborhoods. Since distances obtained from a single point-pattern realization are dependent, uncertainty should be evaluated by a spatial block bootstrap or by parametric simulation rather than by an independent-sample information matrix.

A likelihood-free alternative is minimum contrast. Let
\begin{equation}
\widehat F_n(t)=\frac{1}{n}\sum_{i=1}^{n}\ind\{D_i\leq t\}.
\end{equation}
For a nonnegative weight function $w$ and a maximum reliable distance $\tau$, define
\begin{equation}
(\widehat\lambda,\widehat\mu)
=\argmin_{(\lambda,\mu)\in\mathcal{K}}
\int_0^{\tau}
\left\{\widehat F_n(t)-F_{\mathrm P}(t;\lambda,\mu)\right\}^2
w(t)\dd t,
\label{eq:minimum-contrast}
\end{equation}
where $\mathcal{K}$ is a compact subset of $(0,\infty)^2$. The same construction applies to vertex-origin observations by replacing $F_{\mathrm P}$ with the numerically evaluated $F_{\mathrm I}$.

\begin{proposition}[Consistency of the minimum-contrast estimator]\label{prop:consistency}
Assume that the data-generating parameter $\bm{\eta}_0=(\lambda_0,\mu_0)$ is the unique minimizer over $\mathcal{K}$ of
\begin{equation}
Q(\bm{\eta})=\int_0^{\tau}
\left\{F(t;\bm{\eta}_0)-F(t;\bm{\eta})\right\}^2w(t)\dd t,
\end{equation}
that $F(t;\bm{\eta})$ is continuous in $\bm{\eta}$ uniformly on $[0,\tau]$, and that
\begin{equation}
\sup_{0\leq t\leq\tau}
|\widehat F_n(t)-F(t;\bm{\eta}_0)|\xrightarrow{\Pp}0.
\end{equation}
Then any measurable minimizer of \eqref{eq:minimum-contrast} converges in probability to $\bm{\eta}_0$.
\end{proposition}

\begin{proof}
The stated uniform convergence and compactness imply uniform convergence of the empirical contrast to $Q$. Uniqueness of the population minimizer then gives the result by the standard argmin theorem.
\end{proof}

The uniform convergence assumption can arise from independent replicated windows, increasing-domain ergodic sampling, or a block construction. More general point-process learning and thinning-based cross-validation methods can also be used for tuning and validation \citep{cronie2024pointlearning}.

\subsection{Observed-network estimators and edge handling}

When the line process realization is observed, a direct moment-based calibration is available. For a window $W$ of area $|W|$, let $L_W$ be the total observed centerline length and $N_W$ the number of observed PLCP points on those centerlines. Under the stationary model,
\begin{equation}
\widehat\lambda_{\mathrm{len}}=\frac{L_W}{\pi|W|},
\qquad
\widehat\mu_{\mathrm{line}}=\frac{N_W}{L_W}.
\label{eq:direct-estimators}
\end{equation}
The first estimator follows from \eqref{eq:line-length-density}; conditionally on the line process, the second is the usual Poisson rate estimator.

Finite-window shortest-path distances are right-censored by the line process boundary. A simple guard-zone procedure retains only origins whose distance to the observed boundary exceeds $\tau$. More efficient alternatives can retain censored observations in a composite survival likelihood. In either case, fitted-model assessment should compare the empirical curve with Monte Carlo envelopes produced under the complete observation protocol, including clipping, guard zones, and any map simplification. Systematic discrepancies can reveal anisotropy, nonhomogeneity, line-segment termination, or clustering along individual streets.

\section{Numerical validation and structural trends}\label{sec:numerical}

\paragraph{Numerical integration settings.}
Unless otherwise stated, the one-turn integrals use
$n_\theta=\langle N_\theta\rangle$,
$n_\alpha=\langle N_\alpha\rangle$, and
$n_p=\langle N_p\rangle$ Gauss--Legendre nodes, while the nearest-gateway
integral additionally uses
$n_s=\langle N_s\rangle$ and
$n_\gamma=\langle N_\gamma\rangle$ nodes. As a convergence check, every
order was doubled simultaneously; the largest absolute change over all
displayed CDF values was
$\langle\Delta_{\max}\rangle$. The same settings and domain partitions are
used in the supplied reproducibility code.

\subsection{One-turn laws and typical-intersection bounds}

\begin{figure}[htbp]
    \centering
    \includegraphics[width=0.86\textwidth]{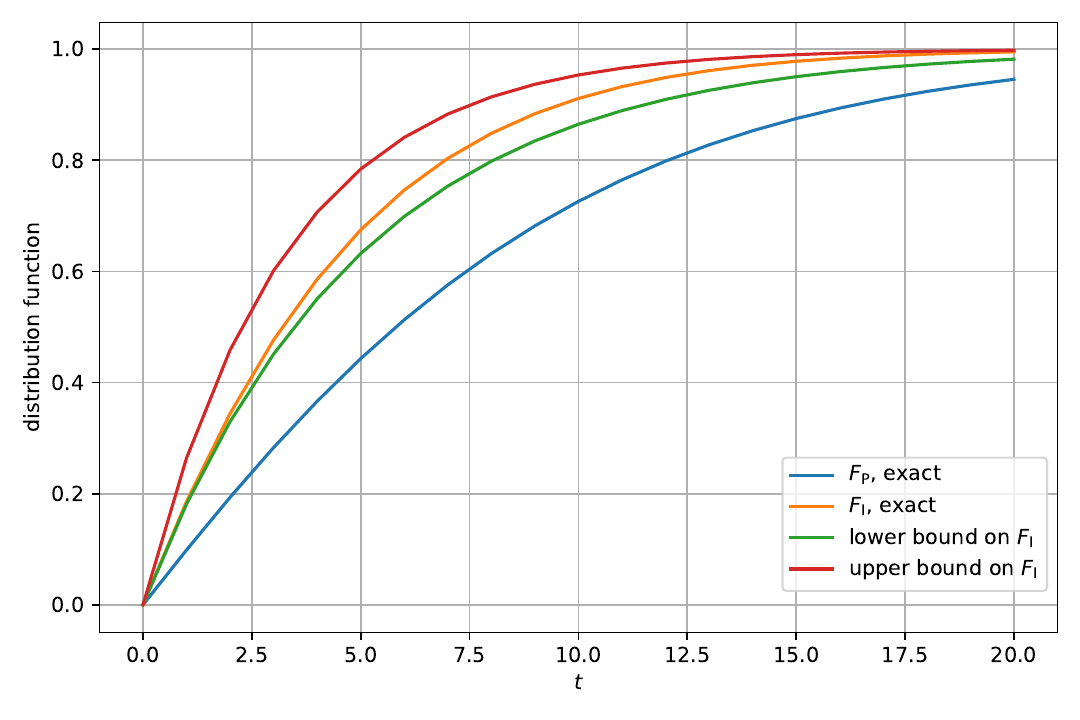}
    \caption{Exact one-turn distributions, typical-intersection bounds, and Monte Carlo validation for $\lambda=0.02$ and $\mu=0.05$.}
    \label{fig:validation}
\end{figure}

We evaluate \eqref{eq:SI} by tensor-product Gauss--Legendre quadrature over $\theta$, $\alpha$, and the bounded offset interval in \eqref{eq:p-support}. Figure~\ref{fig:validation} uses $\lambda=0.02$ and $\mu=0.05$ in inverse-distance units. The markers are Monte Carlo estimates from $60{,}000$ independent typical-intersection constructions at each displayed value of $t$. The simulations draw $\Theta$ from \eqref{eq:angle-density}, generate the residual PLP in the set of offsets that can intersect the two initial segments, and then use the exact union length in \eqref{eq:union-length} for conditional Poisson void sampling.

The quadrature and simulation agree to within Monte Carlo error. The typical-intersection distribution lies above the typical-point distribution because two initial lines, rather than one, are available. The zero-residual-line lower bound is already close when $\mu$ dominates $\lambda$, while the line-exclusion upper bound becomes informative when a substantial part of the gain is caused simply by the presence or absence of nearby intersecting lines. The dependence on physical scale is transparent: replacing $(\lambda,\mu,t)$ by $(\lambda/c,\mu/c,ct)$ leaves all distribution values unchanged.

\subsection{Two-turn sandwich}

\begin{figure}[htbp]
    \centering
    \includegraphics[width=0.86\textwidth]{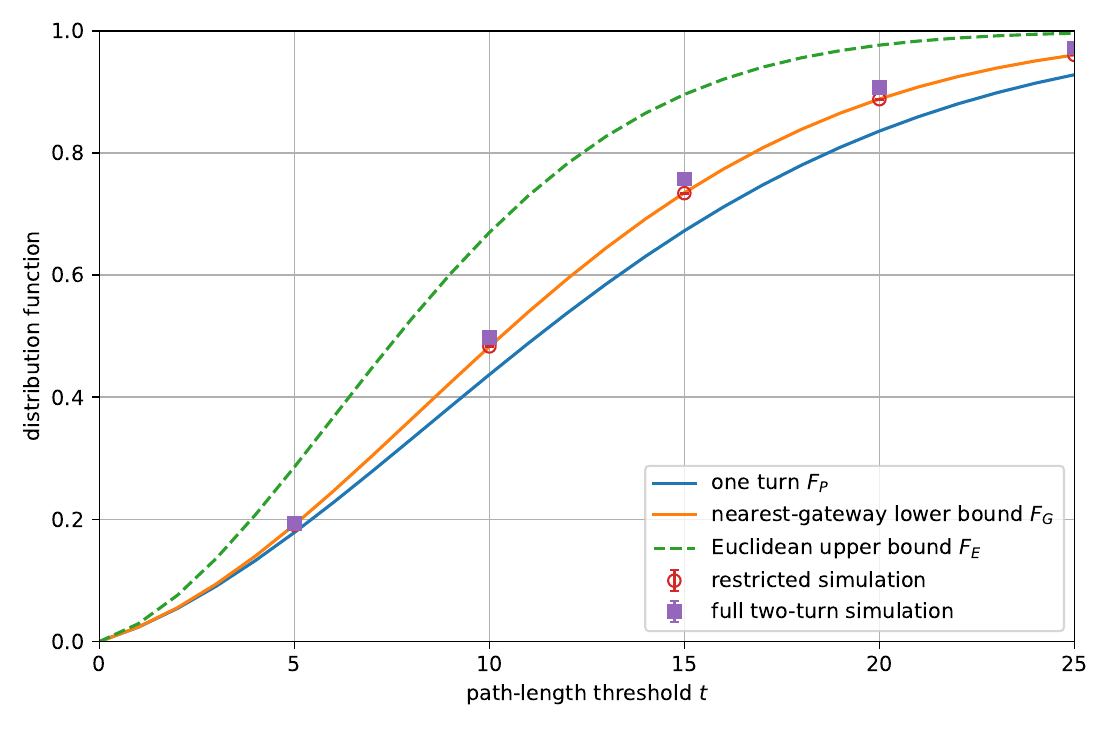}
    \caption{Two-turn bounds and numerical validation for $\lambda=0.1$ and $\mu=0.01$. The analytical nearest-gateway curve is a lower bound on the full two-turn CDF; the Euclidean Palm curve is an upper bound. Error bars show $95\%$ Monte Carlo intervals and are smaller than, or comparable to, the marker size.}
    \label{fig:two-turn-validation}
\end{figure}

Figure~\ref{fig:two-turn-validation} evaluates \Cref{thm:two-turn-sandwich} for $\lambda=0.1$ and $\mu=0.01$ in inverse-distance units, a regime in which additional line transfers are visible because intersections are relatively frequent while points along each line are sparse. The nearest-gateway lower bound $F_{\mathrm G}$ is evaluated by Gauss--Legendre quadrature. Open circles are independent simulations of the same restricted path family and coincide with the analytical curve within Monte Carlo error.

The full two-turn distribution is estimated without generating the Cox points explicitly. For each threshold $t$, all residual PLP lines intersecting the disk $B(\bm{0},t)$ are simulated; this is sufficient because every path of length at most $t$ remains inside that disk. We construct and merge all intervals reachable with at most two turns on every line. If their total length is $L_2(t)$, the conditional survival probability is $\exp\{-\mu L_2(t)\}$. Averaging these weights gives a Rao--Blackwellized Monte Carlo estimator of $F_2(t)$.

At $t=10$, for example, the one-turn CDF is approximately $0.437$, the analytical nearest-gateway lower bound is $0.483$, and the full two-turn simulation is approximately $0.498$. The gateway construction therefore captures about three quarters of the improvement from one to two turns in this regime. Across the displayed thresholds, the ordering in \eqref{eq:two-turn-sandwich} is respected. The Euclidean upper bound is intentionally conservative: it ignores whether a point inside the disk can be reached using only two line transfers, but it supplies a simple rigorous ceiling against which future tighter bounds can be compared.

\section{OpenStreetMap-calibrated urban accessibility}\label{sec:urban}

\subsection{Calibration from a street-network window}

Street-network centerlines can be downloaded and processed with OSMnx \citep{boeing2025osmnx}. Let $W$ be a projected planar window, let $L_W$ be the total undirected street-centerline length after avoiding double counting of bidirectional edges, and let $N_W$ be the number of available vehicles or facilities observed on those centerlines. Equation~\eqref{eq:direct-estimators} gives the model calibration
\begin{equation}
\widehat\lambda=\frac{L_W}{\pi|W|},
\qquad
\widehat\mu=\frac{N_W}{L_W}.
\label{eq:osm-calibration}
\end{equation}
The first identity is a Crofton/Fubini relation and is dimensionally different from an inverse street-density formula. It is important to use a projected coordinate reference system and a centerline representation appropriate to the application.

\begin{figure}[htbp]
    \centering
    \begin{subfigure}[t]{0.47\textwidth}
        \centering
        \includegraphics[width=\linewidth]{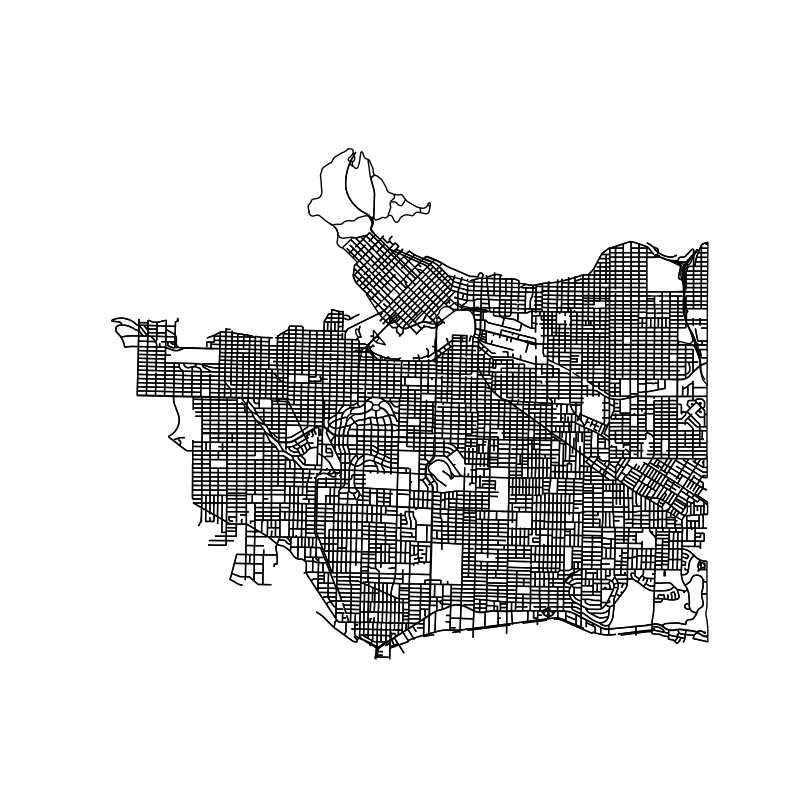}
        \caption{Vancouver.}
    \end{subfigure}\hfill
    \begin{subfigure}[t]{0.47\textwidth}
        \centering
        \includegraphics[width=\linewidth]{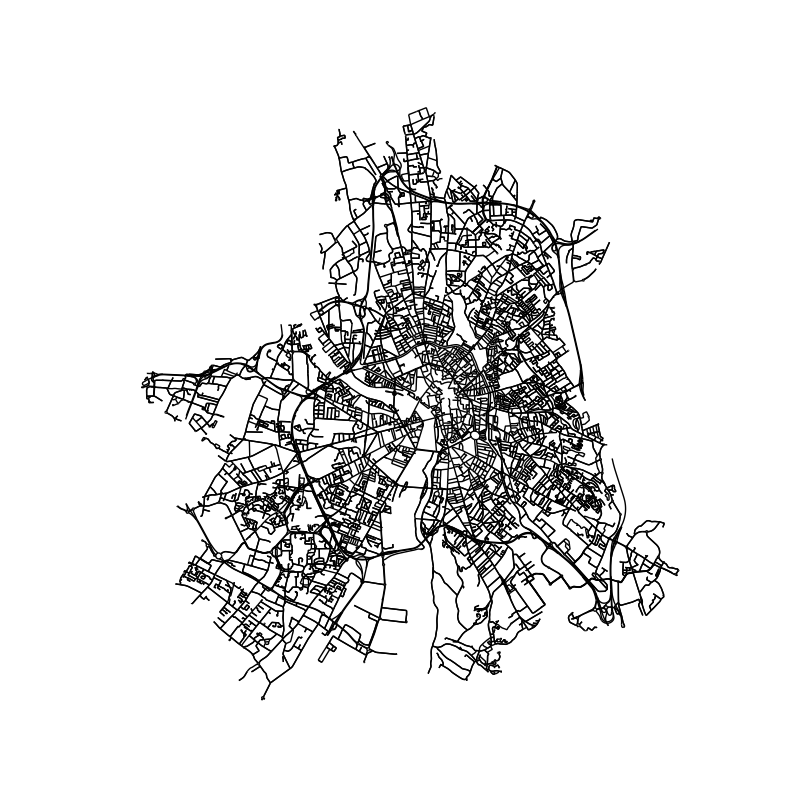}
        \caption{Toulouse.}
    \end{subfigure}
    \caption{Illustrative OSMnx street-network windows used to obtain line-length-density calibrations. The maps are calibration examples rather than a formal goodness-of-fit claim for an isotropic PLP.}
    \label{fig:osm-maps}
\end{figure}

For the illustrative snapshots in \Cref{fig:osm-maps}, the corresponding working Hesse-space intensities are approximately $0.0065\,\mathrm{m}^{-1}$ for Vancouver and $0.0052\,\mathrm{m}^{-1}$ for Toulouse. These values correspond to mean line-length densities $\pi\lambda$ and should be recomputed from the exact extraction windows, filters, and simplification settings used in the final reproducible analysis.

\subsection{Pickup-distance and service-guarantee functionals}

Consider an available-vehicle process with $\mu=0.02\,\mathrm{m}^{-1}$, corresponding to a mean along-street spacing of $50\,\mathrm{m}$. Using the Toulouse working value $\lambda=0.0052\,\mathrm{m}^{-1}$, numerical integration gives
\begin{align}
\E[D_{\mathrm P}]&=\int_0^{\infty}S_{\mathrm P}(t)\dd t\approx20.41\,\mathrm{m},\\
\E[D_{\mathrm I}]&=\int_0^{\infty}S_{\mathrm I}(t)\dd t\approx11.25\,\mathrm{m}.
\label{eq:urban-means}
\end{align}
The corresponding $0.9$ quantiles are approximately $45.01\,\mathrm{m}$ and $25.30\,\mathrm{m}$, respectively. These values are consistent with the direct-line benchmarks: without using any intersecting residual line, the mean nearest distance is $1/(2\mu)=25\,\mathrm{m}$ from a typical point and $1/(4\mu)=12.5\,\mathrm{m}$ from a typical intersection.

For a request at a typical intersection, a service target $\Pp(D_{\mathrm I}\leq d_0)\geq\eta$ is evaluated directly from \eqref{eq:FI}. With $d_0=30\,\mathrm{m}$ and $\eta=0.9$, the same line intensity requires
\begin{equation}
\mu\gtrsim0.01648\,\mathrm{m}^{-1},
\end{equation}
corresponding to a mean along-street spacing of about $60.7\,\mathrm{m}$. If vehicles travel at constant effective speed $v$, the pickup-time distribution follows from $T=D/v$ without further spatial approximation.

For a typical-point request in which two transfers are permitted, \Cref{thm:two-turn-sandwich} gives the simulation-free service bracket
\begin{equation}
F_{\mathrm G}(d_0)\leq\Pp(D_2\leq d_0)\leq F_{\mathrm E}(d_0),
\label{eq:urban-two-turn-bracket}
\end{equation}
and the corresponding mean-distance ordering
\begin{equation}
\E[R]\leq\E[D_2]\leq\E[D_{\mathrm G}]\leq\E[D_{\mathrm P}].
\label{eq:urban-two-turn-means}
\end{equation}
Thus the corrected two-turn result can be used either as an analytic guarantee or as a validation envelope for a city-specific routing simulation.

\subsection{Bias of a matched-intensity planar Poisson surrogate}

A planar PPP with the same areal point intensity $\rho=\pi\lambda\mu$ has Euclidean nearest-neighbor distribution
\begin{equation}
F_{\mathrm{PPP}}(t)=1-\exp(-\pi^2\lambda\mu t^2).
\label{eq:matched-PPP}
\end{equation}
Matching intensity therefore does not match the distance law.

\begin{theorem}[Unique crossing with the planar surrogate]\label{thm:unique-crossing}
Let
\begin{align}
A(t)&=2\mu t+4\lambda t
-\frac{2\lambda}{\mu}\bigl(1-\exp(-2\mu t)\bigr),\\
g(t)&=\pi^2\lambda\mu t^2-A(t).
\end{align}
There exists a unique $t^{\star}>0$ such that
\begin{equation}
F_{\mathrm P}(t)>F_{\mathrm{PPP}}(t)
\quad\text{for }0<t<t^{\star},
\qquad
F_{\mathrm P}(t)<F_{\mathrm{PPP}}(t)
\quad\text{for }t>t^{\star}.
\label{eq:crossing-order}
\end{equation}
\end{theorem}

\begin{proof}
Equality of the two distribution functions is equivalent to $g(t)=0$. We have $g(0)=0$, $g'(0)=-2\mu<0$, and
\begin{equation}
g''(t)=2\lambda\mu\left\{\pi^2-4\exp(-2\mu t)\right\}>0.
\end{equation}
Thus $g$ is strictly convex, initially decreases below zero, and tends to infinity because of its quadratic term. It therefore has exactly one positive root. The sign of $g(t)$ gives \eqref{eq:crossing-order}.
\end{proof}

For the Toulouse illustration above, $t^{\star}\approx50.54\,\mathrm{m}$. Hence a Euclidean matched-intensity model underestimates the probability of very short one-turn path-lengths but eventually has a lighter tail. The two laws are not ordered by first-order stochastic dominance, so a Euclidean correction factor cannot be valid uniformly across service thresholds.

\section{Discussion, limitations, and extensions}\label{sec:discussion}

The exact one-turn laws create local accessibility statistics rather than complete routing models. The nearest-gateway construction extends the analysis to two turns while preserving a rigorous bound direction: it retains one full second-generation branch instead of treating dependent branches as independent. It is most informative when one or two transfers are operationally natural, when intersections are sparse enough that unrestricted detours are costly, or when an analytic benchmark is needed for evaluating numerical routing methods. The unrestricted shortest-path problem remains open for an isotropic PLCP because reachable sets become dependent through repeated intersections and cycles.

The PLP is stationary, isotropic, and composed of infinite lines. Real street networks are finite, often anisotropic, and contain dead ends, curved streets, overpasses, and heterogeneous point intensities. Consequently, \eqref{eq:osm-calibration} should be followed by model checking, not treated as proof that the PLP is an adequate generative description. Orientation histograms can diagnose anisotropy; spatially varying line or point intensities can be introduced through inhomogeneous line processes or line-specific Cox intensities; and finite-segment models can be used when street termination is central. The proposed distance distributions remain useful as parsimonious reference models and as components of simulation-based inference.

A statistically important extension is to estimate the full distribution under dependent origin sampling in one large realization, with edge correction and increasing-domain theory. Other directions include marked points, anisotropic line-orientation measures, spatio-temporal availability, joint nearest-neighbor summaries from several origin types, and cross-validation or prediction-error criteria tailored to random geometric supports. Tighter two-turn bounds could retain the nearest gateway on each side of the Palm point or use several ordered gateways, at the cost of higher-dimensional union-length integrals. Exact or controlled-accuracy multi-turn approximations remain especially valuable, but any feasible-set approximation must preserve the direction of the induced stochastic bound.

\section{Conclusion}\label{sec:conclusion}

We derived network-constrained distance summaries in a Poisson line Cox process from two statistically distinct origins. The one-turn typical-point law is closed form and yields an explicit, identifiable hazard decomposition. The typical-intersection law is an exact bounded integral that correctly averages over the sine-distributed Palm angle and accounts for overlap of the reachable intervals on every residual line. For paths with at most two turns, we replaced an unjustified branch-independence approximation by a nearest-gateway construction with an exact probability-generating-functional law. This yields a rigorous lower bound on the full two-turn CDF, while the reduced-Palm Euclidean nearest-neighbor law yields a rigorous upper bound; Rao--Blackwellized simulation validates the sandwich and quantifies its tightness. The resulting estimators, model-checking workflow, OpenStreetMap calibration, and service-probability brackets turn the distance laws into usable spatial-statistical methodology. Together, the results provide controlled first-order summaries for point processes supported on random linear networks and a rigorous baseline for tighter multi-gateway, anisotropic, nonhomogeneous, and unrestricted-routing extensions.

\section*{Data availability}
The street-network illustrations use publicly available OpenStreetMap data processed with OSMnx. No proprietary data were used. The numerical results in \Cref{sec:numerical,sec:urban} are generated from the equations in the paper and Monte Carlo simulation. Reproducibility code is supplied with the submission.


%

\bibliographystyle{elsarticle-harv}
\bibliography{references/references}

\end{document}